\documentclass[aps,prl,twocolumn,floatfix,showpacs]{revtex4}
\usepackage{amssymb}

\usepackage{epsfig}
\usepackage{amsmath}

\begin{document}

\textbf{Schneider \textit{et al.} Reply:} In our original paper \cite{Schneider09} we demonstrate that the results of
magneto transport in graphite can be fully understood within the frame work of the Slonczewski, Weiss and McClure (SWM)
model. The phase of the oscillatory conductivity $\Delta \sigma \propto cos(2\pi B_f/B-2\pi \gamma + \delta)$
corresponds to the expected phase for massive charge carriers with $\gamma=1/2$.  Our results, together with previously
published work \cite{Soule64,Williamson65,Schroeder68}, disagree with the conclusions of the authors of the preceding
Comment \cite{Luk10}, who find evidence for massless Dirac fermions with a phase $\gamma = 0$ \cite{Luk04,Luk06}.

The origin of this controversy is not the ``improper treatment of experimental results''. To analyze our data we used
exactly the same method as in Ref. \cite{Luk04}, \emph{i.e.} the phase and the frequency were extracted directly from
the Fourier--transformed magnetotransport data. Despite the extremely high quality of our data, in which quantum
oscillations are observed for both majority electrons and holes with orbital quantum number up to almost $N=100$, we
find that it is \emph{simply not possible} to reliably estimate the phase from a $1/B$ versus $N$ plot.

\begin{figure}[b]
\begin{center}
\includegraphics[width= 8.0cm]{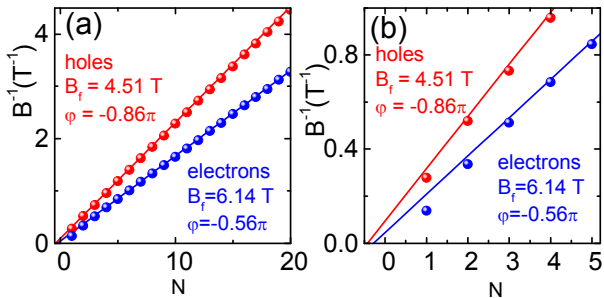}
\end{center}
\caption{(color online) Position of the minima in $\Delta R_{xx}(B)$ as a function of the orbital quantum number for
(a) $N \leq 20$ and (b) $N \leq 5$ corresponding to the range of $N$ used to extract the phase in Ref.~\cite{Luk06} .
The solid lines are the expected linear dependence for the value of $B_f$ and the phase $\varphi$ extracted from the
Fourier transform (Fig.~2 in Ref.~\cite{Schneider09}). A clear deviation from $1/B$ periodicity is seen in (b) for the
$N \leq 2$ electron features.} \label{Fig1}
\end{figure}

Luk'yanchuk and Kopelevich \cite{Luk04,Luk06} analyze a limited number of Shubnikov de Haas oscillations ($1\leq N \leq
5$) at high magnetic field ($B>1$~T). In this region, the electron--hole cross--talk becomes important, leading to the
well documented~\cite{Woollam71a} and considerable movement of the Fermi energy as the quantum limit is approached. The
oscillations are no longer periodic in $1/B$ and the resulting deviation from linearity is clearly seen in our data for
the electron series (Fig.~\ref{Fig1}). Extrapolating data, which is not periodic in $1/B$, to infinite magnetic field
to extract the phase is, at the very least, highly questionable.

A second major problem with Refs.~\cite{Luk04,Luk06}, is that attributing the phase of $\gamma=0$ to Dirac fermions
(holes) at the $H$-point obliges Luk'yanchuk and Kopelevich to invert the well established assignment of the high
frequency series ($B_f=6.14$~T) to electrons at the $K$-point the low frequency series ($B_f=4.51$~T) to holes at the
$H$-point~\cite{Schroeder68}. If correct, this reassignment would have far reaching consequences, changing the position
of the Fermi energy and modifying some of the SWM parameters. A number of problems with Ref.~\cite{Luk04} have already
been pointed out, notably concerning the validity of this reassignment \cite{Mikitik06}. Moreover, the sign of the de
Haas van Alphen signal, invoked by Luk'yanchuk and Kopelevich to lend support to the reassignment, cannot be used to
determine the nature of the charge carriers \cite{Mikitik06}.

Finally, we are somewhat astounded by the suggestion in the comment that their observation of Dirac fermions in
transport signifies the presence of uncoupled layers of graphene. In Ref.~\cite{Luk04} the ``Dirac fermions'' where
assigned to holes at the $H$-point of graphite. While some graphene undoubtedly exists in graphite, it is highly
unlikely that it would dominate the electrical conductivity. We reiterate that the SWM model, which has been
extensively tested using Shubnikov de Haas, de Haas van Alphen, thermopower, magneto-reflectance and Nernst effect
measurements to caliper the Fermi surface of graphite, perfectly predicts the presence of the observed majority
electron and hole pockets~\cite{Soule58,Soule64,Woollam70,Woollam71a,Williamson65,Schroeder68,Schneider09,Zhu10}.

\vspace{0.5cm}
J. M. Schneider, M. Orlita, M. Potemski and D. K. Maude\\
Laboratoire National des Champs Magn\'etiques Intenses, CNRS, 25 avenue des Martyrs, 38042 Grenoble, France.


\begin{thebibliography}{12}
\expandafter\ifx\csname natexlab\endcsname\relax\def\natexlab#1{#1}\fi \expandafter\ifx\csname
bibnamefont\endcsname\relax
  \def\bibnamefont#1{#1}\fi
\expandafter\ifx\csname bibfnamefont\endcsname\relax
  \def\bibfnamefont#1{#1}\fi
\expandafter\ifx\csname citenamefont\endcsname\relax
  \def\citenamefont#1{#1}\fi
\expandafter\ifx\csname url\endcsname\relax
  \def\url#1{\texttt{#1}}\fi
\expandafter\ifx\csname urlprefix\endcsname\relax\def\urlprefix{URL }\fi \providecommand{\bibinfo}[2]{#2}
\providecommand{\eprint}[2][]{\url{#2}}

\bibitem[{\citenamefont{Schneider et~al.}(2009)\citenamefont{Schneider, Orlita,
  Potemski, and Maude}}]{Schneider09}
\bibinfo{author}{\bibfnamefont{J.~M.} \bibnamefont{Schneider}},
  \bibinfo{author}{\bibfnamefont{M.}~\bibnamefont{Orlita}},
  \bibinfo{author}{\bibfnamefont{M.}~\bibnamefont{Potemski}}, \bibnamefont{and}
  \bibinfo{author}{\bibfnamefont{D.~K.} \bibnamefont{Maude}},
  \bibinfo{journal}{Phys. Rev. Lett.} \textbf{\bibinfo{volume}{102}},
  \bibinfo{pages}{166403} (\bibinfo{year}{2009}).

\bibitem[{\citenamefont{Soule et~al.}(1964)\citenamefont{Soule, McClure, and
  Smith}}]{Soule64}
\bibinfo{author}{\bibfnamefont{D.~E.} \bibnamefont{Soule}},
  \bibinfo{author}{\bibfnamefont{J.~W.} \bibnamefont{McClure}},
  \bibnamefont{and} \bibinfo{author}{\bibfnamefont{L.~B.} \bibnamefont{Smith}},
  \bibinfo{journal}{Phys. Rev.} \textbf{\bibinfo{volume}{134}},
  \bibinfo{pages}{A453} (\bibinfo{year}{1964}).

\bibitem[{\citenamefont{Williamson et~al.}(1965)\citenamefont{Williamson,
  Foner, and Dresselhaus}}]{Williamson65}
\bibinfo{author}{\bibfnamefont{S.~J.} \bibnamefont{Williamson}},
  \bibinfo{author}{\bibfnamefont{S.}~\bibnamefont{Foner}}, \bibnamefont{and}
  \bibinfo{author}{\bibfnamefont{M.~S.} \bibnamefont{Dresselhaus}},
  \bibinfo{journal}{Phys. Rev.} \textbf{\bibinfo{volume}{140}},
  \bibinfo{pages}{A1429} (\bibinfo{year}{1965}).

\bibitem[{\citenamefont{Schroeder et~al.}(1968)\citenamefont{Schroeder,
  Dresselhaus, and Javan}}]{Schroeder68}
\bibinfo{author}{\bibfnamefont{P.~R.} \bibnamefont{Schroeder}},
  \bibinfo{author}{\bibfnamefont{M.~S.} \bibnamefont{Dresselhaus}},
  \bibnamefont{and} \bibinfo{author}{\bibfnamefont{A.}~\bibnamefont{Javan}},
  \bibinfo{journal}{Phys. Rev. Lett.} \textbf{\bibinfo{volume}{20}},
  \bibinfo{pages}{1292} (\bibinfo{year}{1968}).

\bibitem[{\citenamefont{Luk'yanchuk and Kopelevich}(2010)}]{Luk10}
\bibinfo{author}{\bibfnamefont{I.~A.} \bibnamefont{Luk'yanchuk}}
  \bibnamefont{and}
  \bibinfo{author}{\bibfnamefont{Y.}~\bibnamefont{Kopelevich}},
  \bibinfo{journal}{preceding {Comment}, Phys. Rev. Lett.}
  (\bibinfo{year}{2010}).

\bibitem[{\citenamefont{Luk'yanchuk and Kopelevich}(2004)}]{Luk04}
\bibinfo{author}{\bibfnamefont{I.~A.} \bibnamefont{Luk'yanchuk}}
  \bibnamefont{and}
  \bibinfo{author}{\bibfnamefont{Y.}~\bibnamefont{Kopelevich}},
  \bibinfo{journal}{Phys. Rev. Lett.} \textbf{\bibinfo{volume}{93}},
  \bibinfo{pages}{166402} (\bibinfo{year}{2004}).

\bibitem[{\citenamefont{Luk'yanchuk and Kopelevich}(2006)}]{Luk06}
\bibinfo{author}{\bibfnamefont{I.~A.} \bibnamefont{Luk'yanchuk}}
  \bibnamefont{and}
  \bibinfo{author}{\bibfnamefont{Y.}~\bibnamefont{Kopelevich}},
  \bibinfo{journal}{Phys. Rev. Lett.} \textbf{\bibinfo{volume}{97}},
  \bibinfo{pages}{256801} (\bibinfo{year}{2006}).

\bibitem[{\citenamefont{Woollam}(1971)}]{Woollam71a}
\bibinfo{author}{\bibfnamefont{J.~A.} \bibnamefont{Woollam}},
  \bibinfo{journal}{Phys. Rev. B} \textbf{\bibinfo{volume}{3}},
  \bibinfo{pages}{1148} (\bibinfo{year}{1971}).

\bibitem[{\citenamefont{Mikitik and {Yu. V. Sharlai}}(2006)}]{Mikitik06}
\bibinfo{author}{\bibfnamefont{G.~P.} \bibnamefont{Mikitik}} \bibnamefont{and}
  \bibinfo{author}{\bibnamefont{{Yu. V. Sharlai}}}, \bibinfo{journal}{Phys.
  Rev. B} \textbf{\bibinfo{volume}{73}}, \bibinfo{pages}{235112}
  (\bibinfo{year}{2006}).

\bibitem[{\citenamefont{Soule}(1958)}]{Soule58}
\bibinfo{author}{\bibfnamefont{D.~E.} \bibnamefont{Soule}},
  \bibinfo{journal}{Phys. Rev.} \textbf{\bibinfo{volume}{112}},
  \bibinfo{pages}{698} (\bibinfo{year}{1958}).

\bibitem[{\citenamefont{Woollam}(1970)}]{Woollam70}
\bibinfo{author}{\bibfnamefont{J.~A.} \bibnamefont{Woollam}},
  \bibinfo{journal}{Phys. Rev. Lett.} \textbf{\bibinfo{volume}{70}},
  \bibinfo{pages}{811} (\bibinfo{year}{1970}).

\bibitem[{\citenamefont{Zhu et~al.}(2010)\citenamefont{Zhu, Yang, Fauqu\'e,
  Kopelevich, and Behnia}}]{Zhu10}
\bibinfo{author}{\bibfnamefont{Z.}~\bibnamefont{Zhu}},
  \bibinfo{author}{\bibfnamefont{H.}~\bibnamefont{Yang}},
  \bibinfo{author}{\bibfnamefont{B.}~\bibnamefont{Fauqu\'e}},
  \bibinfo{author}{\bibfnamefont{Y.}~\bibnamefont{Kopelevich}},
  \bibnamefont{and} \bibinfo{author}{\bibfnamefont{K.}~\bibnamefont{Behnia}},
  \bibinfo{journal}{Nature Physics} \textbf{\bibinfo{volume}{6}},
  \bibinfo{pages}{26} (\bibinfo{year}{2010}).

\end{thebibliography}

\end{document}